\documentclass[prl,preprintnumbers,twocolumn,superscriptaddress, amsmath,amssymb]{revtex4}

\usepackage{graphicx}
\usepackage{dcolumn}
\usepackage{bm}
\usepackage{verbatim}
\usepackage{CJK}
\usepackage{braket}
\usepackage{xcolor}

\begin{document}


\title{Limits on Axion Couplings from the first 80-day data of PandaX-II Experiment}

\def\shKeyLab{INPAC and Department of Physics and Astronomy, Shanghai Jiao Tong University,
Shanghai Laboratory for Particle Physics and Cosmology, Shanghai 200240, China}
\def\pku{School of Physics, Peking University, Beijing 100871, China}
\def\YaLongSD{Yalong River Hydropower Development Company, Ltd., 288 Shuanglin Road, Chengdu 610051, China}
\def\IAP{Shanghai Institute of Applied Physics, Chinese Academy of Sciences, 201800 Shanghai, China}
\def\CHEPpku{Center of High Energy Physics, Peking University, Beijing 100871, China}
\def\SDU{School of Physics and Key Laboratory of Particle Physics and Particle Irradiation (MOE), Shandong University, Jinan 250100, China}
\def\UMD{Department of Physics, University of Maryland, College Park, Maryland 20742, USA}
\def\TDLee{Tsung-Dao Lee Institute, Shanghai 200240, China}
\def\MESJTU{School of Mechanical Engineering, Shanghai Jiao Tong University, Shanghai 200240, China}
\def\THU{Department of Physics, Tsinghua University, Beijing 100084, China}

\author{Changbo Fu}\affiliation{\shKeyLab}
\author{Xiaopeng Zhou}
\email[Corresponding author:]{zhouxp@pku.edu.cn}
\affiliation{\pku}

\author{Xun Chen}\affiliation{\shKeyLab}
\author{Yunhua Chen} \affiliation{\YaLongSD}
\author{Xiangyi Cui} \affiliation{\shKeyLab}
\author{Deqing Fang} \affiliation{\IAP}
\author{Karl Giboni} \affiliation{\shKeyLab}
\author{Franco Giuliani} \affiliation{\shKeyLab}
\author{Ke Han} \affiliation{\shKeyLab}
\author{Xingtao Huang} \affiliation{\SDU}

\author{Xiangdong Ji} \thanks{PandaX Spokesperson, xdji@sjtu.edu.cn}
\affiliation{\shKeyLab}\affiliation{\CHEPpku}\affiliation{\TDLee}

\author{Yonglin Ju} \affiliation{\MESJTU}
\author{Siao Lei} \affiliation{\shKeyLab}
\author{ Shaoli Li} \affiliation{\shKeyLab}
\author{Huaxuan Liu} \affiliation{\MESJTU}
\author{Jianglai Liu} \affiliation{\shKeyLab}\affiliation{\TDLee}
\author{Yugang Ma} \affiliation{\IAP}
\author{Yajun Mao} \affiliation{\pku}
\author{Xiangxiang Ren } \affiliation{\shKeyLab}
\author{Andi Tan} \affiliation{\UMD}
\author{Hongwei Wang} \affiliation{\IAP}
\author{Jimin Wang} \affiliation{\YaLongSD}
\author{Meng Wang} \affiliation{\SDU}
\author{Qiuhong Wang} \affiliation{\IAP}
\author{Siguang Wang} \affiliation{\pku}
\author{Xuming Wang} \affiliation{\shKeyLab}
\author{Zhou Wang} \affiliation{\MESJTU}
\author{Shiyong Wu} \affiliation{\YaLongSD}
\author{Mengjiao Xiao} \affiliation{\UMD}\affiliation{\CHEPpku}
\author{Pengwei Xie} \affiliation{\shKeyLab}
\author{ Binbin Yan} \affiliation{\SDU}
\author{Yong Yang} \affiliation{\shKeyLab}
\author{Jianfeng Yue} \affiliation{\YaLongSD}
\author{Hongguang Zhang} \affiliation{\shKeyLab}
\author{Tao Zhang} \affiliation{\shKeyLab}
\author{Li Zhao} \affiliation{\shKeyLab}
\author{Ning Zhou} \affiliation{\shKeyLab}
\collaboration{PandaX-II Collaboration}
\date{\today}
\begin{abstract}
We report new searches for the solar axions and galactic axion-like dark matter particles,
using the first low-background data from PandaX-II experiment at China Jinping Underground Laboratory, corresponding to a total exposure of about
$2.7\times 10^4$ kg$\cdot$day. No solar axion or galactic axion-like dark matter particle candidate has been identified.
The upper limit on the axion-electron coupling ($g_{Ae}$) from the solar flux
is found to be about $4.35 \times 10^{-12}$ in mass range from $10^{-5}$
to 1~keV/$c^2$ with 90\% confidence level, similar to the recent LUX result. 
We also report a new best limit from the $^{57}$Fe de-excitation. 
On the other hand, the upper limit from the galactic axions is on the
order of $10^{-13}$ in the mass range from 1~keV/$c^2$ to 10~keV/$c^2$ with 90\%
confidence level, slightly improved compared with the LUX.
\end{abstract}
\maketitle
\maketitle


Various theories beyond the Standard Model have predicted new weakly-coupled light $U_{A}(1)$ Goldstone bosons~\cite{HILL1988253, Axion-Review-2010ARNPS, PQ-Axion1977PRL,Gildener1976,Marsh2016},
which may answer many fundamental questions related to CP violation, possible Lorentz violation, dark
matter~\cite{CPT-PhysRev.136.B1542, CP-Axion-RMP2010, Loong-PRL2016,Axion-DM-2001},
etc. The axion, a pseudo-scalar Goldstone boson introduced
by Wilczek~\cite{Wilczek-PhysRevLett.40.279} and
Weinberg~\cite{Wenberg-PhysRevLett.40.223},
arises when the so-called Peccei-Quinn symmetry~\cite{PQ-Axion1977PRL}
in quantum chromodynamics (QCD) is spontaneously broken, which provides 
a natural solution to the so-called ``strong CP problem" in QCD.

Different experimental methods~\cite{annurev2015} have been employed to search for the QCD axion or axion-like particles (ALPs), including  the helioscopes~\cite{Axion-Hilioscopes-2009}, Light Shining through a Wall~\cite{Axion-ALPS-2010}, 
microwave cavities~\cite{Cavity-Axion-2003-RMP.75.777},
nuclear magnetic resonance~\cite{yan2015searching},
and the so-called axioelectrical effect~\cite{Axioelectric-PRD1987}.
Similar to the photoelectric effect, an axioelectrical effect refers to that
an axion or ALP is absorbed by a bound electron in an atom,
producing a free electron emission, i.e:
\begin{equation}
a+e+Z\rightarrow e'+Z \ .
\end{equation}
The cross section for this process is related to
that of the photoelectric effect through~\cite{Axioelectric-PRD2010, Axion-EDELWEISS-2013},
\begin{equation}\label{eq.Ae.CS}
\sigma_{Ae}(E_A)=\sigma_{pe}(E_A)
\frac{g_{Ae}^2}{\beta}
\frac{3 E_A^2}{16\pi\alpha m^2_e}
\left(
1- \frac{\beta^{2/3}}{3}
\right) \ ,
\end{equation}
where $\sigma_{pe}$ is the photoelectric cross section,
$g_{Ae}$ the coupling constant between the axion and electron,
$E_A$  the incident axion energy,
$\alpha$ the fine structure constant, $m_e$ the mass of electron,
and $\beta=v/c$ the axion velocity. 
The recoiling electron kinetic energy is $E_A - E_B$, where $E_B$ is the binding energy of the electron.
Therefore, the recoiling electron signals (ER) in direct dark matter search
experiments can be used to search for
axions or ALPs. Previous reports on the axion couplings from dark matter experiments
can be found in the Ref.
\cite{CDMS, CoGeNT, Axion-EDELWEISS-2013, CDEX-2017, XMASS, Xenon100-2014, Xenon100-2017, LUX2017}.

PandaX, located at China Jinping Underground Laboratory (CJPL),
is a series of experiments utilizing the xenon time-projection-chamber detectors.
The total mass in the target is about 120 kg in
PandaX-I~\cite{PandaX-I-2014-1stResult, PANDA-PRD2015},
and about 580 kg in PandaX-II~\cite{PandaX-PRL-2016, PandaX-2017PRL-Spin}.
By combining the prompt scintillation photons ($S1$)
and the delayed electroluminescence photons ($S2$),
PandaX has excellent ($\sim$cm) vertex reconstruction capabilities, which allow
powerful background suppression via self-shielding and fiducialization. To set the scale,
the ER background rate in PandaX-II has reached a very low level of
2.0$\times$10$^{-3}$~evt/keV/day (=2.0 mDRU), which makes it a highly sensitive detector
to search for axion-electron scattering.
In this paper, we report the new constraints on axion/ALP electron coupling
strength $g_{Ae}$ by using the first low-background data in PandaX-II experiment
(Run 9) with a total exposure of about $2.7 \times 10^4$ kg$\cdot$day,
one of the largest reported xenon data sets in the world to date.

\begin{figure}[htbp]
\begin{center}
  \includegraphics[width=8.5cm]{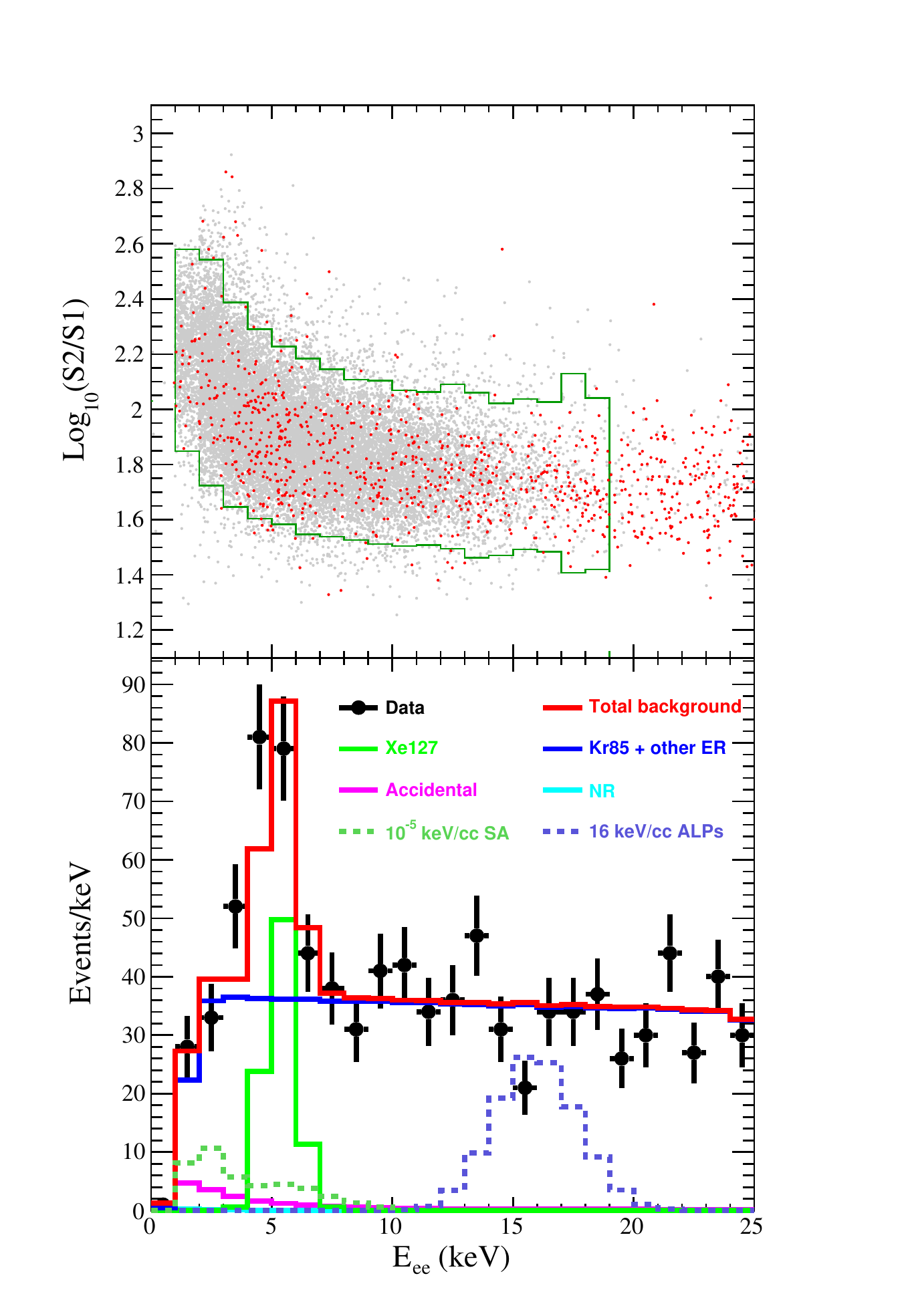}
\caption{
Upper: Event distribution obtained in $\log_{10}(S2/S1)$ vs. $E_{ee}$
in PandaX-II experiment;
The $\pm2\sigma$ contours for CH$_3$T calibration data is indicated as the green box,
and the dark matter data are drawn as red crosses.
Lower: the combined energy spectrum with data (histogram with uncertainties)
compared to the best fit (red histogram), with
individual background components indicated (see Ref.~\cite{PandaX-PRL-2016}).
We also plot here the estimated $10^{-5}$~keV/$c^2$ solar axion and 16~keV/$c^2$ ALPs 
spectra assuming that $g_{Ae}$ equals $5\times10^{-12}$ and $5\times10^{-13}$ respectively.
See text for details.
}
\label{fig.S1S2}
\end{center}
\end{figure}

As in Ref.~\cite{PandaX-PRL-2016}, the Run 9 data was divided into 14 time bins
according to the temporal change of detector parameters and background rates.
For each event, the electron-equivalent energy $E_{ee}$ was reconstructed from $S1$
and $S2$ as
\begin{equation}\label{eq.combE}
E_{ee} = \frac{S1}{\text{PDE}}+\frac{S2}{\text{EEE}\times{\text{SEG}}}\,,
\end{equation}
where PDE, EEE, and SEG are photon detection efficiency, electron extraction
efficiency, and single electron gain, respectively.
Most of the data cuts were identical to those in
Ref.~\cite{PandaX-PRL-2016,PandaX-2017PRL-Spin},
except we enlarged the energy window of search by replacing the upper $S1$ and $S2$ cuts with a single cut of $E_{ee} < 25$~keV. Based on the tritiated methane (CH$_3$T)  calibration, the detection threshold was determined to be 1.29~keV, and in high energy region the detection efficiency was 94\%. In total, 942 candidate events survived. 
The distribution of these events in log$_{10}(S2/S1)$ vs. reconstructed energy is 
shown in the upper panel in Fig.~\ref{fig.S1S2} as the red dots.
For comparison, the distribution bands corresponding to the ER calibration data
from the tritium with a $\beta$-decay end point at 18.6~keV
is overlaid in the figure (shadow dots). The physical data are largely consistent with ER events. The measured combined energy spectrum is shown in the lower panel of
Fig.~\ref{fig.S1S2}. In the energy range shown,
the ER background is dominated by $^{85}$Kr (flat) and $^{127}$Xe (peak around 5 keV).

The solar axions may be produced through the following processes~\cite{Axion-EDELWEISS-2013}: Compton-like scattering (C), axion-Bremsstrahlung (B), atomic-recombination (R),
and atomic-deexcitation(D). Given $g_{Ae}$, they can all be calculated.

\begin{figure}[h]
\begin{center}
\includegraphics[width=8.5cm]{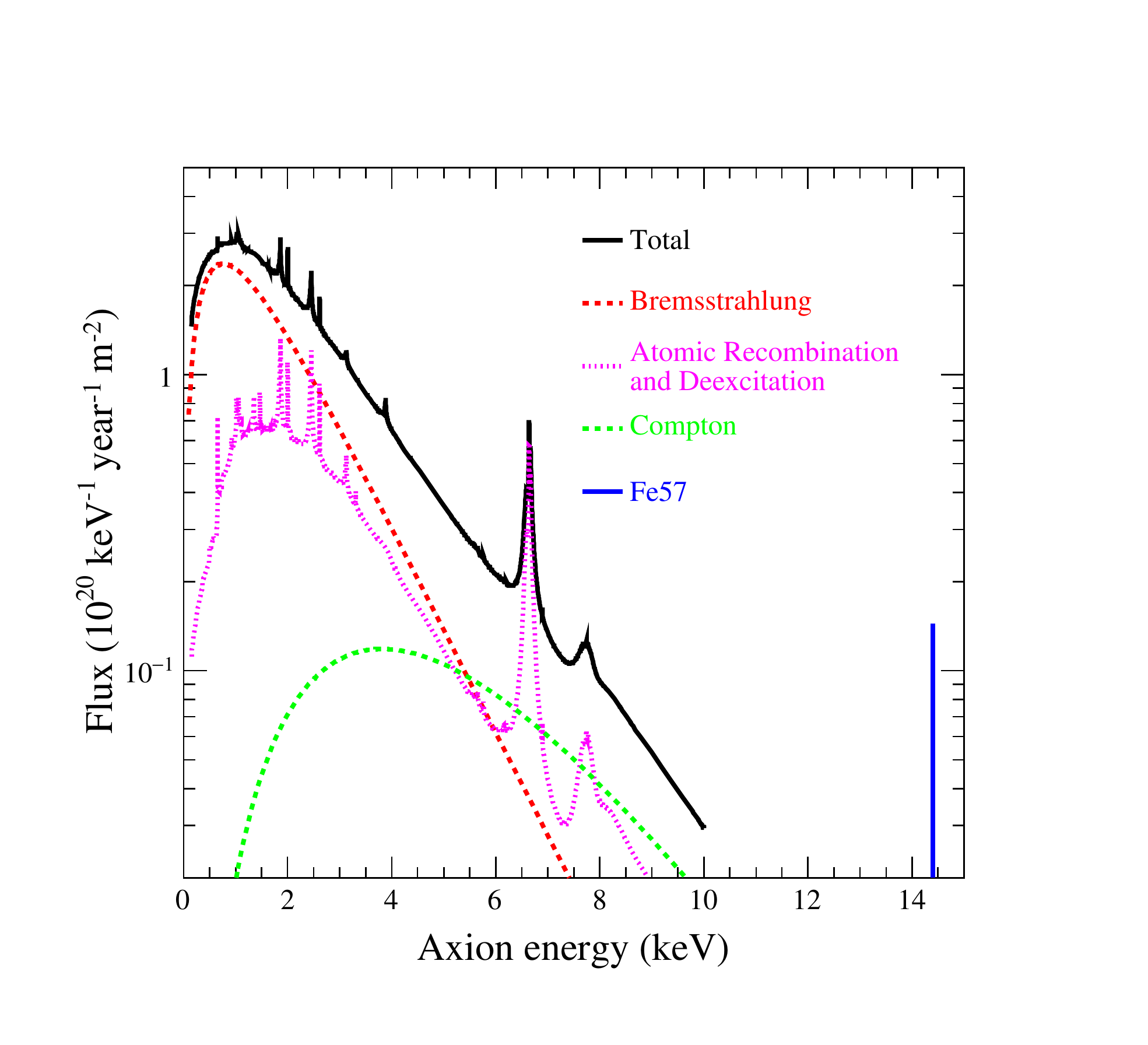}
\caption{
The expected solar-axion flux at the earth's orbit deduced from theoretical models~\cite{Axion-EDELWEISS-2013}.
Five mechanisms are considered here:
Compton-like scattering (C),
axion-bremsstrahlung (B),
atomic-recombination (R),
and atomic-deexcitation (D).
See the text for details.
The 14.4 keV line is generated by  $^{57}{\rm Fe}^*$ deexcitation.
In this plot, the corresponding axion parameters
are set to be $g_{Ae} = 10^{-13}$ and $g_{AN}^{eff}= 10^{-8}$.
}
\label{fig.flux}
\end{center}
\end{figure}

We took the calculations from Ref.~\cite{SA-Flux-2013JCAP} 
as our input axion spectrum for the axion energy range of $E_A<10$ keV,
which is valid for an axion mass less than~1 keV/$c^2$.
As shown in Fig.~\ref{fig.flux}, towards the lower energy (1--2 keV), the flux is
dominated by axion-Bremsstrahlung process, 
and at the high energy region (9--10 keV),  by Compton-like scattering.

Additionally, deexcitation of $^{57}{\rm Fe}^*$ may also generated
monoenergetic axions, i.e.
$^{57}{\rm Fe}^*\rightarrow\,^{57}{\rm Fe}+a+14.4$~keV~\cite{Fe57-Axion-PRL1995},
This monoenergetic axion flux at the Earth's orbit was estimated to be~\cite{Axion-cast-2009,Axion-EDELWEISS-2013}:
\begin{equation}\label{eq.flux.Fe56}
\Phi_{14.4}
=4.56\times10^{23} \cdot (g_{AN}^{\rm{eff}})^2
\left(
\frac{k_A}{k_\gamma}
\right)^3
\ \rm{cm}^{-2} \rm{s}^{-1}
\end{equation}
where $k_A$/$k_\gamma$ is the momentum ratio between the axion and the gamma,
and the $g_{AN}^{\rm{eff}}$ is a model and axion mass-dependent coupling constant between the axion and nucleus. In this work, we took the benchmark function of $g_{AN}^{\rm{eff}}$ in the so-called Dine-Fischler-Srednicki-Zhitnitskii (DFSZ) model as in Ref.~\cite{Axion-EDELWEISS-2013}.

The axion (or ALP) flux from the Milky Way dark matter (MWDM) halo can be
estimated as follows.
The MWDM density at the Earth location is
$\rho_{\rm DM}^{(E)}\simeq 0.3$ GeV/cm$^3$~\cite{DM-density-2012}.
If all the MWDM is composed of ALPs,
the corresponding ALP flux $\Phi_{A}$ can then be written as
\begin{equation}
\Phi_{A}=\rho_{\rm DM}^{(E)} \cdot v_A/m_A= 9\times10^{15}\frac{\beta}{m_A},
\end{equation}
where $v_A$ is the axion velocity relative to the Earth,
$m_A$ is the axion mass in unit keV/$c^2$, and $\beta=v_A/c$.
Considering the same axion electron scattering mechanism,
the expected ALP detection rate $R$ can be expressed as~\cite{Det-Rate-PRD2008}:
\begin{equation}
R\simeq g^2_{Ae}\left( \frac{1.2\times 10^{19}}{A} \right)
\left( \frac{m_A}{ {\rm keV}/c^2} \right)
\left( \frac{\sigma_{pe}}{\rm barn} \right)
{\text{kg}}^{-1}{\text{day}}^{-1},
\end{equation}
where $A=131.9$ is the average mass number of the xenon.


PandaX-II data can be fitted by combining the axion signal and background models.
The axion or ALP signals are computed by combining incident fluxes above with
the axion-electron scattering cross section in Eq.~(\ref{eq.Ae.CS}).  The background
estimates are identical to those in Ref.~\cite{PandaX-PRL-2016}, including
$^{127}$Xe, $^{85}$Kr and other ER background, accidental, and nuclear recoil (NR)
backgrounds. As in Ref.~\cite{PandaX-PRL-2016},
Geant4-based~\cite{Geant4-2003} simulation using NEST~\cite{NEST}
ER and NR models, together with the efficiencies in $S1$ and $S2$, produce the
signal and background probability distribution functions in $S1$ and $S2$. For illustration,
an example axion or ALP signal is overlaid in the
lower panel of Fig.~\ref{fig.S1S2}. For each pair of values of axion mass and
$g_{Ae}$, profile likelihood ratio statistic~\cite{Likelihood-PhysRevD.84.052003} is
constructed. 
The likelihood function~\cite{PandaX-PRL-2016} used here is
\begin{equation}
\label{eq:likelihood}
  \mathcal{L}_{\rm pandax} = \big[\prod_{n=1}^{\textrm{bins}}\mathcal{L}_{n}\big] \times \big[\rm Gauss(\delta_{\rm A}, \sigma_{\rm A}) \prod_{b}\rm Gauss(\delta_b, \sigma_b)\big]\,,
\end{equation}
where
\begin{eqnarray}
  \mathcal{L}_{n} = &&{\rm Poisson}(N_{\rm m}^n|N_{\rm ept}^n)\times \\\nonumber
  &&\Bigg[\displaystyle{\prod_{i=1}^{N_{\rm m}^n}}\left(\frac{N_{\rm A}^n(1+\delta_{\rm A})P_{\rm A}^n(S1^i,S2^i)}{N_{\rm ept}^n}\right. \\\nonumber
&&+ \left.\displaystyle\sum_b \frac{N_{b}^n(1+\delta_{b})P_{b}^n(S1^i,S2^i)}{N_{\rm ept}^n}\right)\Bigg]\,.
\end{eqnarray}
$N_{m}$ is the event number measured experimentally, and $N_{ept}$ is the expected event number. Axion (or ALPs) and background numbers are represented as $N_{A}$ and $N_{b}$. Their probability distribution functions (PDFs), $P_{A}$ and $P_{b}$, are generated using NEST-based models. Here background was divided in five independent components, $^{127}$Xe, $^{85}$Kr, other ER, accidental coincidence and neutron. $\sigma$ and $\delta$ are systematic uncertainties and nuisance parameters for individual components with values listed in Ref.~\cite{PandaX-PRL-2016}. 

\begin{figure}[h]
\begin{center}
\includegraphics[width=0.5\textwidth]{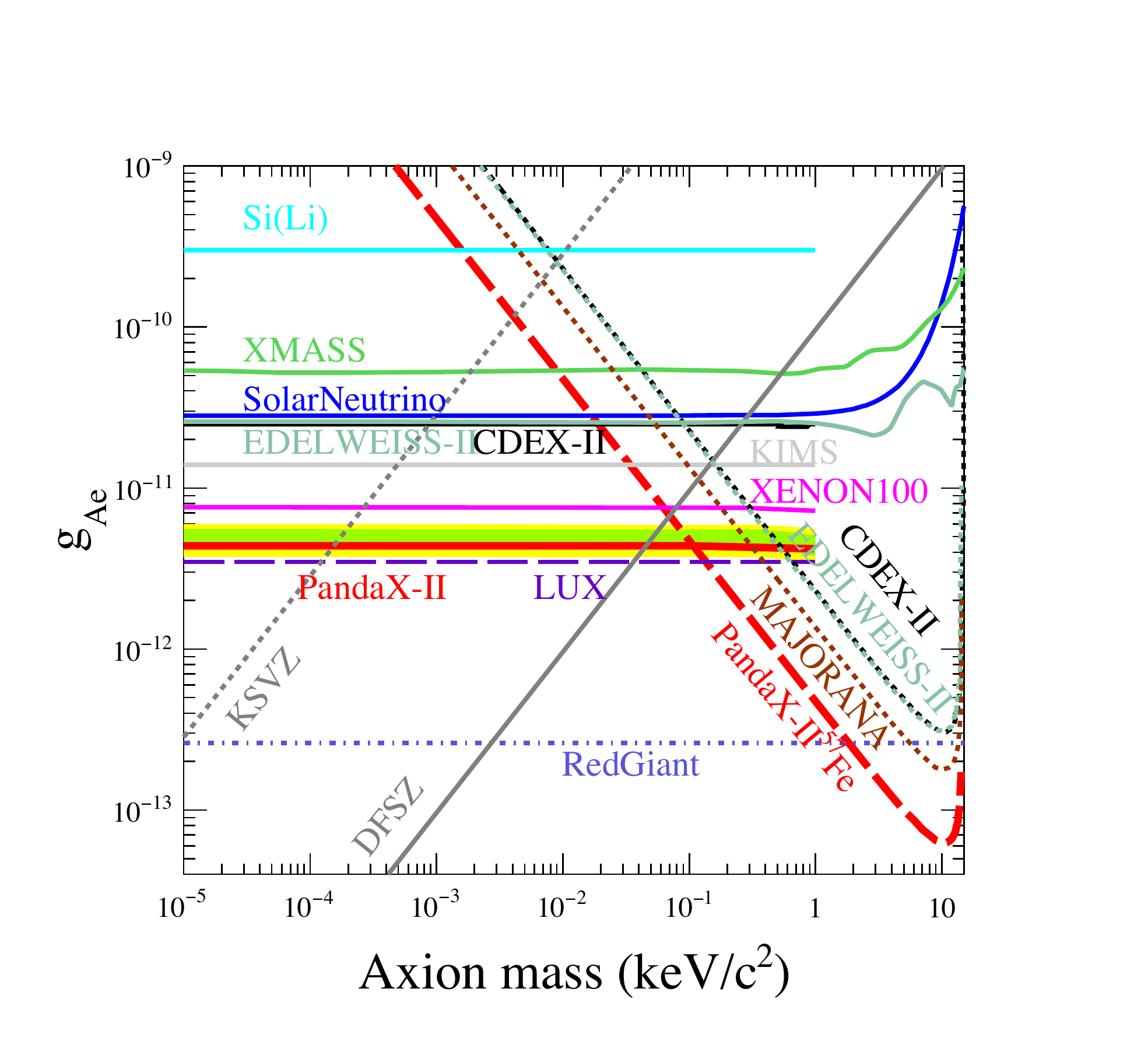}
\caption{
The 90\% upper limits on solar CBRD axion (solid red line) and
14.4 keV $^{57}$Fe solar axion in $g_{Ae}$ vs. $m_A$.
The constrains from other representative experiments are also shown,
including those using solar neutrinos~\cite{SolarNeutrino}, and data from
Si(Li) target~\cite{SiLi},
CDEX-II~\cite{CDEX-2017},
XMASS~\cite{XMASS},
EDELWEISS-II~\cite{Axion-EDELWEISS-2013}, KIMS~\cite{KIMS2016}, 
XENON100~\cite{Xenon100-2014}, LUX~\cite{LUX2017}, M\textsc{ajorana} D\textsc{emonstrator}~\cite{MAJ2017} (converted to $g_{Ae}$ using the same benchmark DSFZ model values for $g_{AN}^{eff}$ as in this paper), and observations of Red Giant~\cite{RedGiant}.
The benchmarks of the QCD axion models,
DFSZ~\cite{DFSZ1981simple,Axion-EDELWEISS-2013}
and KSVZ (Kim-Shifman-Vainstein-Zakharov)~\cite{KSVZ-1980NPB,Axion-EDELWEISS-2013},
are also displayed.
}
\label{SA.Result}
\end{center}
\end{figure}

For all channels we considered, data are consistent with no axion signals. 
For the solar axion from the CBRD mechanisms shown above, 
the results are presented in Fig.~\ref{SA.Result},
in which the 90\% confidence level (CL) is shown as the red solid curve.
The upper limit of $g_{Ae}$ is set to about $g_{Ae}\le4\times 10^{-13}$ with $90$\% CL
in the axion mass range of $10^{-5}<m_A<1$ keV/$c^2$, similar to the recent limit from the 
LUX experiment~\cite{LUX2017}.
Due to the high temperature in the solar core, the axion flux is generally 
independent of its mass, and the axioelectrical  cross section picks up 
a gentle $\beta$-dependence (Eq.~(\ref{eq.Ae.CS})) only when $m_A$ gets closer 
1 keV/$c^2$. Therefore, this limit is largely independent of the axion mass. 
The constraint from the $^{57}$Fe 14.4 keV axion is drawn as red dotted line.
The most sensitive upper limit on $g_{Ae}$ is set at $6\times 10^{-14}$ 
at $m_A=$10 keV/$c^2$, which represents the best such limit to date. 
The fast decline of sensitivity for lower and higher mass 
is primarily due to the linear mass dependence of 
g$_{AN}^{\rm{eff}}$ in the benchmark DSFZ
 model~\cite{DFSZ1981simple}, and the axion momentum 
dependence in Eq.~(\ref{eq.flux.Fe56}), respectively.

\begin{figure}[htbp]
\begin{center}
\includegraphics[width=0.5\textwidth]{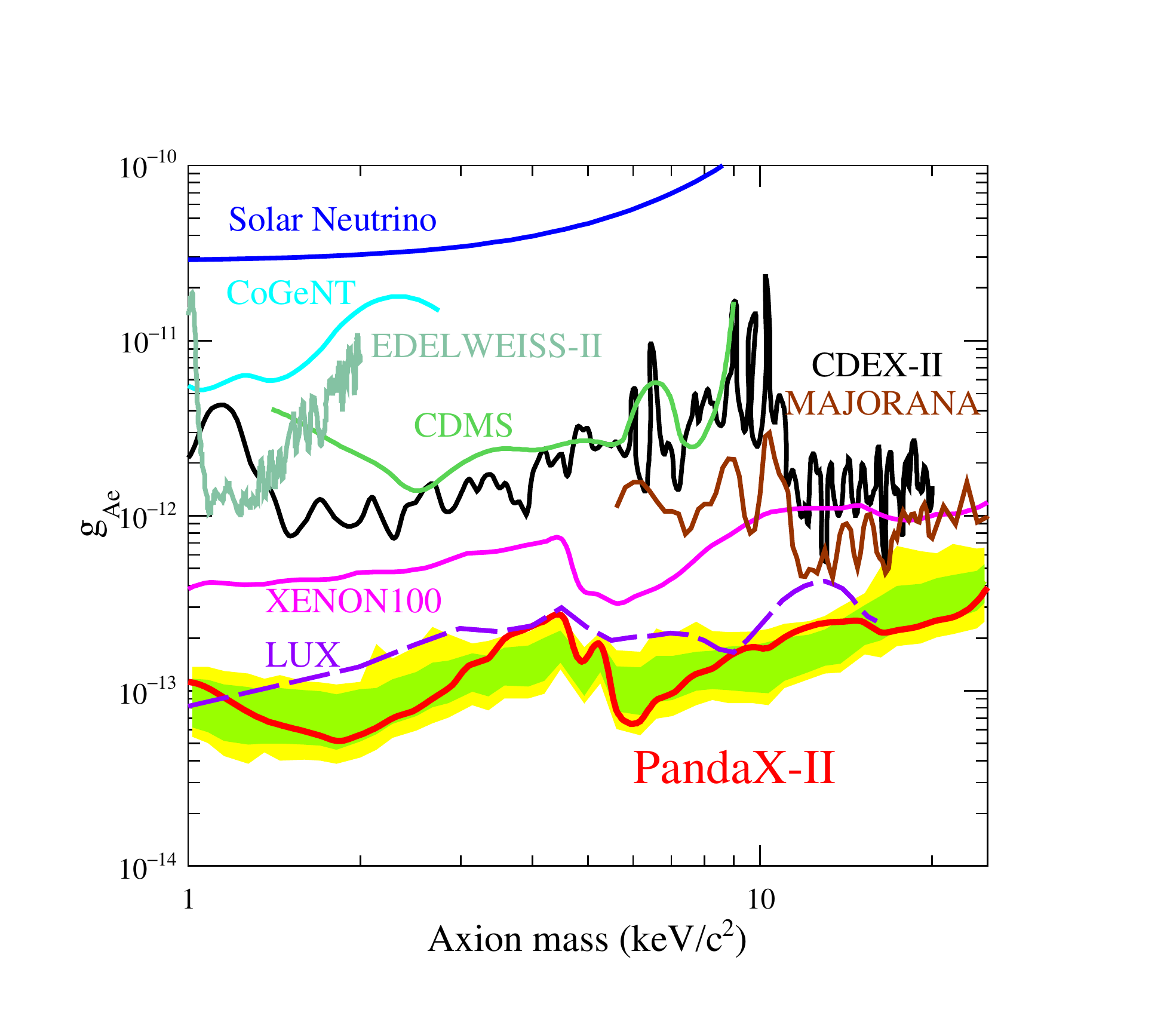}
\caption{
Constraints on $g_{Ae}$ as a function of MWDM ALP mass. PandaX's 90\% limit
is shown as the red curve, with $\pm1\sigma$ and $\pm2\sigma$ sensitivity bands
in green and yellow respectively.
The constraints from other representative experiments are also shown,
including
those from the solar neutrinos~\cite{SolarNeutrino},
data from CDEX-II~\cite{CDEX-2017},
CoGeNT~\cite{CoGeNT},
CDMS~\cite{CDMS},
EDELWEISS-II~\cite{Axion-EDELWEISS-2013}, XENON100~\cite{Xenon100-2017}, LUX~\cite{LUX2017} and M\textsc{ajorana} D\textsc{emonstrator}~\cite{MAJ2017}.}
\label{fig.GA.result}
\end{center}
\end{figure}

The limits on the galactic ALPs are shown in Fig.~\ref{fig.GA.result}.  The 90\% limit to
$g_{Ae}$ is set to be about $\le4\times 10^{-13}$ in the mass range $1<m_A<25$~keV/$c^2$.
This limit is about 3--10 times improved
from the results from XENON100, CDEX-II, and M\textsc{ajorana} D\textsc{emonstrator}~\cite{Xenon100-2017,CDEX-2017,MAJ2017}, and slightly
improved from the LUX's recent result~\cite{LUX2017}. The slightly weakened limit
between 4--6 keV/$c^2$ is due to the $^{127}$Xe background in our detector, as shown in the 
lower panel of Fig.~\ref{fig.S1S2}.




In summary, using the first low-background dark matter search data from
PandaX-II experiment and via the axioelectrical effects,
we have set new limits on the axion-electron coupling
constant $g_{Ae}$ for solar axions and galactic ALPs.
For the solar axions, the limit $g_{Ae}$ is $4.35\times10^{-11}$
for axion mass between $10^{-5}$ to 1 keV/$c^2$, similar to the recent limits from
LUX~\cite{LUX2017}. 
Best limit on $g_{Ae}$ from $^{57}$Fe axion is also reported, 
with the lowest exclusion limit of 6$\times$10$^{-14}$ at a mass of 10 keV/$c^2$. 
For the galactic ALPs, $g_{Ae}$ is constrained to be $<4.3\times10^{-14}$ (90\% C.L.)
for an axion mass between 1 to 25 keV/$c^2$, which
represents the strongest constraints to date. PandaX-II will continue taking data, and
more sensitive search of axion is expected in the future.

\begin{acknowledgments}
This project has been supported by a 985-III grant from Shanghai Jiao Tong University, grants from National Science Foundation of China (Nos. 11365022, 11435008, 11455001, 11505112 and 11525522), and a grant from the Ministry of Science and Technology of China (No. 2016YFA0400301). We thank the support of grants from the Office of Science and Technology, Shanghai Municipal Government (No. 11DZ2260700, No. 16DZ2260200), and the support from the Key Laboratory for Particle Physics, Astrophysics and Cosmology, Ministry of Education. This work is supported in part by the Chinese Academy of Sciences Center for Excellence in Particle Physics (CCEPP) and Hongwen Foundation in Hong Kong. Finally, we thank the following organizations for indispensable logistics and other supports: the CJPL administration and the Yalong River Hydropower Development Company Ltd.
\end{acknowledgments}

\end{document}